\documentclass[11pt,a4paper]{article}

\usepackage{graphicx}

\usepackage{amsmath}
\usepackage{verbatim}
\usepackage{amssymb}
\usepackage{hyperref}
\usepackage[margin=1in]{geometry}

\usepackage{amsthm}
\usepackage{geometry}
\usepackage{float}

\usepackage{setspace}
\spacing{1.3}
\usepackage{fancyhdr}

\usepackage{color} 

\usepackage{mathabx}
\usepackage{overpic}

\usepackage[all]{xy}

\graphicspath{{pics/}}

\newcommand{\al}{\alpha}

\newcommand{\vphi}{\varphi}

\newcommand{\be}{\beta}

\newcommand{\ga}{\gamma}

\newcommand{\de}{\delta}

\newcommand{\om}{\omega}

\newcommand{\na}{\nabla}

\newcommand{\NA}{\nabla}

\newcommand{\bs}{\boldsymbol}

\newcommand{\ra}{\rightarrow}

\newcommand{\lra}{\longrightarrow}

\newcommand{\Ra}{\Rightarrow}

\newcommand{\xra}{\xrightarrow}

\newcommand{\xlra}{\xlongrightarrow}

\newcommand{\rgl}{\rangle}

\newcommand{\lgl}{\langle}

\newcommand{\dash}{\textrm{-}}

\newcommand{\ot}{\otimes}

\newcommand{\bpf}{\begin{proof}}

\newcommand{\epf}{\end{proof}}

\newcommand{\bthm}{\begin{thm}}

\newcommand{\ethm}{\end{thm}}

\newcommand{\bprop}{\begin{prop}}

\newcommand{\eprop}{\end{prop}}

\newcommand{\bcor}{\begin{cor}}

\newcommand{\ecor}{\end{cor}}

\newcommand{\blem}{\begin{lem}}

\newcommand{\elem}{\end{lem}}

\newcommand{\bdefn}{\begin{defn}}

\newcommand{\edefn}{\end{defn}}

\newcommand{\bexmp}{\begin{exmp}}

\newcommand{\eexmp}{\end{exmp}}

\newcommand{\brem}{\begin{rem}}

\newcommand{\erem}{\end{rem}}

\newcommand{\bdia}{\begin{displaymath}\xymatrix}

\newcommand{\edia}{\end{displaymath}}

\newcommand{\beq}{\begin{equation*}\begin{aligned}}

\newcommand{\eeq}{\end{aligned}\end{equation*}}

\newcommand{\bref}{\textbf{Ref}}

\newcommand{\intg}{\mathbb{Z}}

\newcommand{\real}{\mathbb{R}}

\newcommand{\comp}{\mathbb{C}}

\newcommand{\quot}{\mathbb{H}}

\newcommand{\afv}{\mathbb{A}}

\newcommand{\prv}{\mathbb{P}}

\newcommand{\mco}{\mathcal{O}}

\newcommand{\mcc}{\mathcal{C}}

\newcommand{\mcf}{\mathcal{F}}

\newcommand{\mcg}{\mathcal{G}}

\newcommand{\mcs}{\mathcal{S}}

\newcommand{\cp}{\mathbb{CP}}

\newcommand{\mfo}{\mathfrak{O}}

\newcommand{\mfg}{\mathfrak{g}}

\newcommand{\msa}{\mathscr{A}}

\newcommand{\msr}{\mathscr{R}}

\newcommand{\msg}{\mathscr{G}}

\newcommand{\msd}{\mathscr{D}}

\newcommand{\itbf}{\item\textbf}

\newcommand{\seqa}{a_1,...,a_}

\newcommand{\seqx}{x_1,...,x_}

\newcommand{\seqy}{y_1,...,y_}

\newcommand{\seqf}{f_1,...,f_}

\newcommand{\cred}{\textcolor{red}}

\newcommand{\cblue}{\textcolor{blue}}

\newcommand{\mfa}{\mathfrak{a}}

\newcommand{\mfb}{\mathfrak{b}}

\newcommand{\mfm}{\mathfrak{m}}

\newcommand{\mfn}{\mathfrak{n}}

\newcommand{\mfp}{\mathfrak{p}}

\newcommand{\Af}{A_{(f)}}

\newcommand{\shm}{\underline{\rm SHM}}

\newtheorem{thm}{\textbf {Theorem}}[section]

\newtheorem{cor}[thm]{\textbf{Corollary}}

\newtheorem{prop}[thm]{\textbf{Proposition}}

\newtheorem{lem}[thm]{\textbf{Lemma}}

\theoremstyle{definition}

\newtheorem{defn}[thm]{\textbf{Definition}}

\newtheorem{exmp}[thm]{Example}

\theoremstyle{remark}

\newtheorem{rem}[thm]{Remark}

\def\cok{\operatorname{Coker}}

\newcommand{\txi}{\tilde{\xi}}

\newcommand{\bxi}{\bar{\xi}}
\usepackage{hyperref}
\usepackage{comment}
\usepackage{float}

\newcommand{\bz}{\bar{z}}

\DeclareMathOperator{\tr}{trunk}

\author{Nithin Kavi \\ \textit{Harvard College} \\ \textit{Email: nithinkavi@college.harvard.edu}} 

\title{MapReduce for Counting Word Frequencies with MPI and GPUs}
\date{}
\def\allfiles{}
\begin{document}
\bibliographystyle{plain}
\maketitle
\thispagestyle{empty}

\begin{abstract}
\thispagestyle{empty}


In this project, the goal was to use the Julia programming language and parallelization to write a fast map reduce algorithm to count word frequencies across large numbers of documents. We first implement the word frequency counter algorithm on a CPU using two processes with MPI. Then, we create another implementation, but on a GPU using the Julia CUDA library, though not using the in built map reduce algorithm within FoldsCUDA.jl. After doing this, we apply our CPU and GPU algorithms to count the frequencies of words in speeches given by Presidents George W Bush, Barack H Obama, Donald J Trump, and Joseph R Biden with the aim of finding patterns in word choice that could be used to uniquely identify each President. We find that each President does have certain words that they use distinctly more often than their fellow Presidents, and these words are not surprising given the political climate at the time.

The goal of this project was to create faster MapReduce algorithms in Julia on the CPU and GPU than the ones that have already been written previously. We present some simple cases of mapping functions where our GPU algorithm outperforms Julia's FoldsCUDA implementation. We also discuss ideas for further optimizations in the case of counting word frequencies in documents and for these specific mapping functions.  


\end{abstract}


\ifx\allfiles\undefined

\documentclass[12pt,a4paper]{article}


\usepackage{graphicx}

\usepackage{amsmath}

\usepackage{amssymb}
\DeclareMathOperator{\tr}{trunk}
\usepackage{amsthm}

\usepackage{geometry}

\usepackage{fancyhdr}

\usepackage{color} 









\newcommand{\al}{\alpha}

\newcommand{\vphi}{\varphi}

\newcommand{\be}{\beta}

\newcommand{\ga}{\gamma}

\newcommand{\de}{\delta}

\newcommand{\om}{\omega}

\newcommand{\na}{\nabla}

\newcommand{\NA}{\nabla}

\newcommand{\bs}{\boldsymbol}

\newcommand{\ra}{\rightarrow}

\newcommand{\lra}{\longrightarrow}

\newcommand{\Ra}{\Rightarrow}

\newcommand{\xra}{\xrightarrow}

\newcommand{\xlra}{\xlongrightarrow}

\newcommand{\rgl}{\rangle}

\newcommand{\lgl}{\langle}

\newcommand{\dash}{\textrm{-}}

\newcommand{\ot}{\otimes}

\newcommand{\bpf}{\begin{proof}}

\newcommand{\epf}{\end{proof}}

\newcommand{\bthm}{\begin{thm}}

\newcommand{\ethm}{\end{thm}}

\newcommand{\bprop}{\begin{prop}}

\newcommand{\eprop}{\end{prop}}

\newcommand{\bcor}{\begin{cor}}

\newcommand{\ecor}{\end{cor}}

\newcommand{\blem}{\begin{lem}}

\newcommand{\elem}{\end{lem}}

\newcommand{\bdefn}{\begin{defn}}

\newcommand{\edefn}{\end{defn}}

\newcommand{\bexmp}{\begin{exmp}}

\newcommand{\eexmp}{\end{exmp}}

\newcommand{\brem}{\begin{rem}}

\newcommand{\erem}{\end{rem}}

\newcommand{\bdia}{\begin{displaymath}\xymatrix}

\newcommand{\edia}{\end{displaymath}}

\newcommand{\beq}{\begin{equation*}\begin{aligned}}

\newcommand{\eeq}{\end{aligned}\end{equation*}}

\newcommand{\bref}{\textbf{Ref}}

\newcommand{\intg}{\mathbb{Z}}

\newcommand{\real}{\mathbb{R}}

\newcommand{\comp}{\mathbb{C}}

\newcommand{\quot}{\mathbb{H}}

\newcommand{\afv}{\mathbb{A}}

\newcommand{\prv}{\mathbb{P}}

\newcommand{\mco}{\mathcal{O}}

\newcommand{\mcc}{\mathcal{C}}

\newcommand{\mcf}{\mathcal{F}}

\newcommand{\mcg}{\mathcal{G}}

\newcommand{\mcs}{\mathcal{S}}

\newcommand{\cp}{\mathbb{CP}}

\newcommand{\mfo}{\mathfrak{O}}

\newcommand{\mfg}{\mathfrak{g}}

\newcommand{\msa}{\mathscr{A}}

\newcommand{\msr}{\mathscr{R}}

\newcommand{\msg}{\mathscr{G}}

\newcommand{\msd}{\mathscr{D}}

\newcommand{\itbf}{\item\textbf}

\newcommand{\seqa}{a_1,...,a_}

\newcommand{\seqx}{x_1,...,x_}

\newcommand{\seqy}{y_1,...,y_}

\newcommand{\seqf}{f_1,...,f_}

\newcommand{\cred}{\textcolor{red}}

\newcommand{\cblue}{\textcolor{blue}}

\newcommand{\mfa}{\mathfrak{a}}

\newcommand{\mfb}{\mathfrak{b}}

\newcommand{\mfm}{\mathfrak{m}}

\newcommand{\mfn}{\mathfrak{n}}

\newcommand{\mfp}{\mathfrak{p}}

\newcommand{\Af}{A_{(f)}}


\newtheorem{thm}{\textbf {Theorem}}[section]

\newtheorem{cor}[thm]{\textbf{Corollary}}

\newtheorem{prop}[thm]{\textbf{Proposition}}

\newtheorem{lem}[thm]{\textbf{Lemma}}

\newtheorem{conj}[thm]{Conjecture}

\newtheorem{conv}[thm]{Convention}

\newtheorem{prob}[thm]{Problem}

\newtheorem{exer}[thm]{Exercise}

\newtheorem{quest}[thm]{Question}

\theoremstyle{definition}

\newtheorem{defn}[thm]{\textbf{Definition}}

\newtheorem{defns}[thm]{Definitions}

\newtheorem{exmp}[thm]{Example}

\newtheorem{exmps}[thm]{Examples}

\newtheorem{var}[thm]{Variant}

\newtheorem{vars}[thm]{Variants}

\newtheorem{con}[thm]{Construction}

\newtheorem{notn}[thm]{Notation}

\newtheorem{notns}[thm]{Notations}

\theoremstyle{remark}

\newtheorem{rem}[thm]{Remark}

\newtheorem{rems}[thm]{Remarks}

\newtheorem{warn}[thm]{Warning}

\newtheorem{sch}[thm]{Scholium}

\newtheorem{expl}[thm]{Explanations}

\newtheorem*{theorem}{\textbf{Theorem}}

\newtheorem*{corollary}{\textbf{Corollary}}

\newtheorem*{proposition}{\textbf{Proposition}}

\newtheorem*{lemma}{\textbf{Lemma}}

\newtheorem*{example}{\textbf{Example}}

\def\cok{\operatorname{Coker}}

\newcommand{\txi}{\tilde{\xi}}

\newcommand{\bxi}{\bar{\xi}}

\newcommand{\bz}{\bar{z}}



\begin{document}

\bibliographystyle{plain}

\else

\fi

\newpage

\section{Introduction}

MapReduce was first introduced by Google in 2004 in \cite{googlemapreduce} for the purpose of efficiently processing and generating large data sets on a cluster of machines in parallel. Since then, it has been used and updated extensively, and one of the most common implementations is in Hadoop.

MapReduce has been used in many different applications, but one of the canonical examples of how to use it is counting the number of appearances of each word in a set of documents. For a small number of short documents, there is little need for parallelism or MapReduce, and some for loops and a HashMap/Dictionary would easily suffice. However, when there are many long documents, the need to read them in parallel becomes much more important.

The goal of this project is to use Julia and parallel computing to write a faster MapReduce algorithm for the word frequency in documents problem. In Section 2, we explore an implementation of MapReduce using MPI for multiprocessing, which we find to not be optimal. Then, in Section 3, we switch to an implementation on the GPU. We also utilize parallelism for the map phase, though that does not generalize as easily to the GPU as we would like. We include a time analysis to study the efficiencies of each step of MapReduce and how the timings vary with and without parallelization on the CPU and how the timing changes when we use a GPU. We find that all the small additions required in the reduce stage are actually much faster on the GPU when there are a large number of long documents.

After the map reduce algorithm is written and optimized, we switch to testing it in Section 4. We compare our GPU implementation of MapReduce to the GPU implementation of reduce in the FoldsCUDA library. The FoldsCUDA library implementation of MapReduce is a very general implementation that is applicable to a variety of map functions and reduce operations. We study the specific cases of adding up the square roots of a large set of numbers and of computing the alternating harmonic sum of a set of numbers, and we see that with an elementary GPU kernel, we can outperform the FoldsCUDA implementation by significant multiplicative factors.

Then, we use our word frequency implemention towards counting words in sets of documents in Section 4.3. The documents we use are speeches from Presidents George W Bush, Barack H Obama, Donald J Trump, and Joseph R Biden, and we see what words each US President uses most often. In order to approximately control for the types of speeches given, we include each President's Inauguration speeches and State of the Union speeches from their first term. We find that the use of common English words makes it difficult to differentiate US Presidents from each other based on their most common words (the most common word for all four Presidents was ``the" which is not surprising and also not helpful in distinguishing them); however, we find that there are less common words that each President uses given the political climate and their party which does differentiate them from each other. 

All code referenced in this paper can be accessed at https://github.com/thinkinavi24/Julia-MapReduce-Implementations. In Section 5, we discuss future research and optimizations, including ideas for a faster GPU implementation and other possible documents to study.

\newpage 

\ifx\allfiles\undefined

\documentclass[12pt,a4paper]{article}


\usepackage{graphicx}

\usepackage{amsmath}

\usepackage{amssymb}
\DeclareMathOperator{\tr}{trunk}
\usepackage{amsthm}

\usepackage{geometry}

\usepackage{fancyhdr}

\usepackage{color} 

\usepackage{mathabx}

\usepackage{overpic}









\newcommand{\shm}{\underline{\rm SHM}}

\newcommand{\al}{\alpha}

\newcommand{\vphi}{\varphi}

\newcommand{\be}{\beta}

\newcommand{\ga}{\gamma}

\newcommand{\de}{\delta}

\newcommand{\om}{\omega}

\newcommand{\na}{\nabla}

\newcommand{\NA}{\nabla}

\newcommand{\bs}{\boldsymbol}

\newcommand{\ra}{\rightarrow}

\newcommand{\lra}{\longrightarrow}

\newcommand{\Ra}{\Rightarrow}

\newcommand{\xra}{\xrightarrow}

\newcommand{\xlra}{\xlongrightarrow}

\newcommand{\rgl}{\rangle}

\newcommand{\lgl}{\langle}

\newcommand{\dash}{\textrm{-}}

\newcommand{\ot}{\otimes}

\newcommand{\bpf}{\begin{proof}}

\newcommand{\epf}{\end{proof}}

\newcommand{\bthm}{\begin{thm}}

\newcommand{\ethm}{\end{thm}}

\newcommand{\bprop}{\begin{prop}}

\newcommand{\eprop}{\end{prop}}

\newcommand{\bcor}{\begin{cor}}

\newcommand{\ecor}{\end{cor}}

\newcommand{\blem}{\begin{lem}}

\newcommand{\elem}{\end{lem}}

\newcommand{\bdefn}{\begin{defn}}

\newcommand{\edefn}{\end{defn}}

\newcommand{\bexmp}{\begin{exmp}}

\newcommand{\eexmp}{\end{exmp}}

\newcommand{\brem}{\begin{rem}}

\newcommand{\erem}{\end{rem}}

\newcommand{\bdia}{\begin{displaymath}\xymatrix}

\newcommand{\edia}{\end{displaymath}}

\newcommand{\beq}{\begin{equation*}\begin{aligned}}

\newcommand{\eeq}{\end{aligned}\end{equation*}}

\newcommand{\bref}{\textbf{Ref}}

\newcommand{\intg}{\mathbb{Z}}

\newcommand{\real}{\mathbb{R}}

\newcommand{\comp}{\mathbb{C}}

\newcommand{\quot}{\mathbb{H}}

\newcommand{\afv}{\mathbb{A}}

\newcommand{\prv}{\mathbb{P}}

\newcommand{\mco}{\mathcal{O}}

\newcommand{\mcc}{\mathcal{C}}

\newcommand{\mcf}{\mathcal{F}}

\newcommand{\mcg}{\mathcal{G}}

\newcommand{\mcs}{\mathcal{S}}

\newcommand{\cp}{\mathbb{CP}}

\newcommand{\mfo}{\mathfrak{O}}

\newcommand{\mfg}{\mathfrak{g}}

\newcommand{\msa}{\mathscr{A}}

\newcommand{\msr}{\mathscr{R}}

\newcommand{\msg}{\mathscr{G}}

\newcommand{\msd}{\mathscr{D}}

\newcommand{\itbf}{\item\textbf}

\newcommand{\seqa}{a_1,...,a_}

\newcommand{\seqx}{x_1,...,x_}

\newcommand{\seqy}{y_1,...,y_}

\newcommand{\seqf}{f_1,...,f_}

\newcommand{\cred}{\textcolor{red}}

\newcommand{\cblue}{\textcolor{blue}}

\newcommand{\mfa}{\mathfrak{a}}

\newcommand{\mfb}{\mathfrak{b}}

\newcommand{\mfm}{\mathfrak{m}}

\newcommand{\mfn}{\mathfrak{n}}

\newcommand{\mfp}{\mathfrak{p}}

\newcommand{\Af}{A_{(f)}}


\newtheorem{thm}{\textbf {Theorem}}[section]

\newtheorem{cor}[thm]{\textbf{Corollary}}

\newtheorem{prop}[thm]{\textbf{Proposition}}

\newtheorem{lem}[thm]{\textbf{Lemma}}

\newtheorem{conj}[thm]{Conjecture}

\newtheorem{conv}[thm]{Convention}

\newtheorem{prob}[thm]{Problem}

\newtheorem{exer}[thm]{Exercise}

\newtheorem{quest}[thm]{Question}

\theoremstyle{definition}

\newtheorem{defn}[thm]{\textbf{Definition}}

\newtheorem{defns}[thm]{Definitions}

\newtheorem{exmp}[thm]{Example}

\newtheorem{exmps}[thm]{Examples}

\newtheorem{var}[thm]{Variant}

\newtheorem{vars}[thm]{Variants}

\newtheorem{con}[thm]{Construction}

\newtheorem{notn}[thm]{Notation}

\newtheorem{notns}[thm]{Notations}

\theoremstyle{remark}

\newtheorem{rem}[thm]{Remark}

\newtheorem{rems}[thm]{Remarks}

\newtheorem{warn}[thm]{Warning}

\newtheorem{sch}[thm]{Scholium}

\newtheorem{expl}[thm]{Explanations}

\newtheorem*{theorem}{\textbf{Theorem}}

\newtheorem*{corollary}{\textbf{Corollary}}

\newtheorem*{proposition}{\textbf{Proposition}}

\newtheorem*{lemma}{\textbf{Lemma}}

\newtheorem*{example}{\textbf{Example}}

\def\cok{\operatorname{Coker}}

\newcommand{\txi}{\tilde{\xi}}

\newcommand{\bxi}{\bar{\xi}}

\newcommand{\bz}{\bar{z}}

\begin{document}

\bibliographystyle{plain}

\else

\fi
\section{MPI Implementation}

We first begin with an MPI Implementation that has two processes. While ideally we would generalize this to $n$ processes for any $n$, we find that this MPI Implementation does not generalize easily as described in more detail below, motivating the use of GPUs as explored in Section 3. All the code and results that are referenced in this section can be found in the file titled MPI\_MapReduce\_TwoProcesses.jl.

The MPI Implementation aims to execute both the map stage and the reduce stage in parallel. The key idea behind it is to first split the documents and then sort the words in each of the documents. We also remove all punctuation and convert to lowercase in order to allow the algorithm to recognize that ``Dog" and ``dog." would be the same word. Since we are looking to eventually reduce along pairs with the same keys, in this case the keys being the words, we want to group similar words together. The proxy for this is to sort all of the words in alphabetical order and let one process handle the first half of the words and the other process handle the second half of the words. By doing this, we ensure that most of the key value pairs with the same keys are handled by the same process. The caveat is that there may be a word where some of its occurrences are handled by one process and its other occurrences are handled by the other process. Thus there is not complete reduction.

For example, if document 1 is ``I want to test MapReduce" and document 2 is ``MapReduce is a cool algorithm to test.", then we get: 

(``cool" $\to$ 1, ``algorithm" $\to$ 1, ``is" $\to$ 1, ``mapreduce" $\to$ 1, ``a" $\to$ 1, ``i" $\to$ 1), (``want" $\to$ 1, ``test" $\to$ 2, ``mapreduce" $\to$ 1, ``to" $\to$ 2).

Note that the word ``mapreduce" is not fully reduced. This is occurs because ``mapreduce" is near the middle alphabetically of all the words across both documents (counting duplicates of words, so we do not mean the union of unique elements, but rather the set of all words).

In order to process the words in this way, we assign separate documents to each of the two processes, and each process sorts the words in its own documents. Then, it sends half of its words to the other process, and the other process does the same. This requires several steps in MPI:

\begin{enumerate}
    \item The number of words being sent must be itself sent, so that the other process can allocate an array with that amount of memory to receive the words.
    \item Each of the words must be converted into a Char Array because MPI will not send strings because they are not isbitstype.
    \item The lengths of each of the words must be sent so that the other process knows how large an array to allocate for each of the words.
    \item The actual words must be sent over as Char Arrays. One process sends over its top half of words alphabetically, while the other process sends over its bottom half of words alphabetically. This ensures that each process at the end has roughly the first half of all the words in alphabetical order and the second process has the other half of all the words in alphabetical order, which means that most duplicates will be on the same process.
    \item On the other process, the char arrays must be converted back into words.
    \item Steps 1-4 must occur for both processes, and the ids of each of the sends/receives must match up so that MPI knows which messages to receive where.
\end{enumerate}

The two process case was intended to be the precursor for a more general $n$ process case. The algorithm for the more general $n$ process case would be as follows:

\begin{enumerate}
    \item If there are $k_j$ total words in the documents on process $j$, then this process must send $\frac{k_j}{n}$ to each of the other $n - 1$ processes.
    \item Each of the words must be converted into a Char Array because MPI will not send strings because they are not isbitstype.
    \item The actual words must be sent over as Char Arrays. Process $j$ should send over all of the words not part of the $\frac{j}{n}$ alphabetical section (in other words, divide up all the words on process $j$ into $n$ equal sets, and send over all words not in the $j$th set, where the words in the $k$th set are sent to process $k$).  This ensures that process $j$ will have the words that are roughly in the $j/n$ portion of all of the words in alphabetical order, which means that most duplicates will be on the same process.
    \item On the other $n - 1$ processes, the char arrays must be converted back into words.
    \item Steps 1-4 must occur for all $n$ processes, and the ids of each of the sends/receives must match up so that MPI knows which messages to receive where 
\end{enumerate}

However, this algorithm does not generalize well for multiple reasons. First, the fact that MPI does not handle strings well requires the conversion to a char array and back, which is evidently not optimal. But the larger issue is that while the alphabetically sorted words means that most duplicates will end up being reduced on the same process, not all will. As the number of processes increases very high (since when processing many long documents having only two or three processes will not be enough to process them quickly), it becomes clear that there will potentially be up to $n -1 $ non-reduced words, which does not seem practical. It is possible to go through and identify the non reduced words in serial, but that seems to defeat the point of parallelism. 

However, we do not believe that this means that MPI cannot be used to implement MapReduce at all, and people have tried implementations previously, such as in \cite{mpimapreduce}, where the authors realized a 25 percent speed up in a 127 note cluster. Rather, this was an attempt to parallelize every aspect of the MapReduce algorithm where $n$ processes would lead to roughly cutting the amount of time by $n$, but it runs into these sorts of complications. Also, while in this project we were focused on writing a MapReduce algorithm specifically geared towards counting duplicate words in documents, it is also worth considering how the algorithm may generalize to the more general case of MapReduce, and the idea of sorting the words alphabetically and processing them in this way does not appear to generalize well to many contexts. In addition, we will discuss a GPU implementation in the next section, and if one has access to multiple GPUs, then MPI could potentially be used to allow the GPUs to communicate with each other and increase the amount of parallelism. We discuss that possibility in Section 5.2: Future Research.

\ifx\allfiles\undefined
\end{document}

\fi

\newpage 

\ifx\allfiles\undefined

\documentclass[12pt,a4paper]{article}


\usepackage{graphicx}

\usepackage{amsmath}

\usepackage{amssymb}
\DeclareMathOperator{\tr}{trunk}
\usepackage{amsthm}

\usepackage{geometry}

\usepackage{fancyhdr}

\usepackage{color} 









\newcommand{\al}{\alpha}

\newcommand{\vphi}{\varphi}

\newcommand{\be}{\beta}

\newcommand{\ga}{\gamma}

\newcommand{\de}{\delta}

\newcommand{\om}{\omega}

\newcommand{\na}{\nabla}

\newcommand{\NA}{\nabla}

\newcommand{\bs}{\boldsymbol}

\newcommand{\ra}{\rightarrow}

\newcommand{\lra}{\longrightarrow}

\newcommand{\Ra}{\Rightarrow}

\newcommand{\xra}{\xrightarrow}

\newcommand{\xlra}{\xlongrightarrow}

\newcommand{\rgl}{\rangle}

\newcommand{\lgl}{\langle}

\newcommand{\dash}{\textrm{-}}

\newcommand{\ot}{\otimes}

\newcommand{\bpf}{\begin{proof}}

\newcommand{\epf}{\end{proof}}

\newcommand{\bthm}{\begin{thm}}

\newcommand{\ethm}{\end{thm}}

\newcommand{\bprop}{\begin{prop}}

\newcommand{\eprop}{\end{prop}}

\newcommand{\bcor}{\begin{cor}}

\newcommand{\ecor}{\end{cor}}

\newcommand{\blem}{\begin{lem}}

\newcommand{\elem}{\end{lem}}

\newcommand{\bdefn}{\begin{defn}}

\newcommand{\edefn}{\end{defn}}

\newcommand{\bexmp}{\begin{exmp}}

\newcommand{\eexmp}{\end{exmp}}

\newcommand{\brem}{\begin{rem}}

\newcommand{\erem}{\end{rem}}

\newcommand{\bdia}{\begin{displaymath}\xymatrix}

\newcommand{\edia}{\end{displaymath}}

\newcommand{\beq}{\begin{equation*}\begin{aligned}}

\newcommand{\eeq}{\end{aligned}\end{equation*}}

\newcommand{\bref}{\textbf{Ref}}

\newcommand{\intg}{\mathbb{Z}}

\newcommand{\real}{\mathbb{R}}

\newcommand{\comp}{\mathbb{C}}

\newcommand{\quot}{\mathbb{H}}

\newcommand{\afv}{\mathbb{A}}

\newcommand{\prv}{\mathbb{P}}

\newcommand{\mco}{\mathcal{O}}

\newcommand{\mcc}{\mathcal{C}}

\newcommand{\mcf}{\mathcal{F}}

\newcommand{\mcg}{\mathcal{G}}

\newcommand{\mcs}{\mathcal{S}}

\newcommand{\cp}{\mathbb{CP}}

\newcommand{\mfo}{\mathfrak{O}}

\newcommand{\mfg}{\mathfrak{g}}

\newcommand{\msa}{\mathscr{A}}

\newcommand{\msr}{\mathscr{R}}

\newcommand{\msg}{\mathscr{G}}

\newcommand{\msd}{\mathscr{D}}

\newcommand{\itbf}{\item\textbf}

\newcommand{\seqa}{a_1,...,a_}

\newcommand{\seqx}{x_1,...,x_}

\newcommand{\seqy}{y_1,...,y_}

\newcommand{\seqf}{f_1,...,f_}

\newcommand{\cred}{\textcolor{red}}

\newcommand{\cblue}{\textcolor{blue}}

\newcommand{\mfa}{\mathfrak{a}}

\newcommand{\mfb}{\mathfrak{b}}

\newcommand{\mfm}{\mathfrak{m}}

\newcommand{\mfn}{\mathfrak{n}}

\newcommand{\mfp}{\mathfrak{p}}

\newcommand{\Af}{A_{(f)}}


\newtheorem{thm}{\textbf {Theorem}}[section]

\newtheorem{cor}[thm]{\textbf{Corollary}}

\newtheorem{prop}[thm]{\textbf{Proposition}}

\newtheorem{lem}[thm]{\textbf{Lemma}}

\newtheorem{conj}[thm]{Conjecture}

\newtheorem{conv}[thm]{Convention}

\newtheorem{prob}[thm]{Problem}

\newtheorem{exer}[thm]{Exercise}

\newtheorem{quest}[thm]{Question}

\theoremstyle{definition}

\newtheorem{defn}[thm]{\textbf{Definition}}

\newtheorem{defns}[thm]{Definitions}

\newtheorem{exmp}[thm]{Example}

\newtheorem{exmps}[thm]{Examples}

\newtheorem{var}[thm]{Variant}

\newtheorem{vars}[thm]{Variants}

\newtheorem{con}[thm]{Construction}

\newtheorem{notn}[thm]{Notation}

\newtheorem{notns}[thm]{Notations}

\theoremstyle{remark}

\newtheorem{rem}[thm]{Remark}

\newtheorem{rems}[thm]{Remarks}

\newtheorem{warn}[thm]{Warning}

\newtheorem{sch}[thm]{Scholium}

\newtheorem{expl}[thm]{Explanations}

\newtheorem*{theorem}{\textbf{Theorem}}

\newtheorem*{corollary}{\textbf{Corollary}}

\newtheorem*{proposition}{\textbf{Proposition}}

\newtheorem*{lemma}{\textbf{Lemma}}

\newtheorem*{example}{\textbf{Example}}

\def\cok{\operatorname{Coker}}

\newcommand{\txi}{\tilde{\xi}}

\newcommand{\bxi}{\bar{\xi}}

\newcommand{\bz}{\bar{z}}

\begin{document}

\bibliographystyle{plain}

\else

\fi

\section{GPU Implementation}

The embarrassingly parallel nature of MapReduce motivates an implementation using the thousands of cores on a GPU. However, there are several obstacles that need to be overcome in order to do this. One of the main ones is that the CUDA library in Julia has no Dictionary or Hashmap implementation available, making counting duplicates of words very tricky. 

It is possible to ``approximate" a Dictionary through an array of keys and an array of values. However, this also has issues on the GPU. First, GPU kernels in Julia do not allow the push! command, which allows for pushing keys and values onto an array (or CuArray) that is growing. This is because the push! command is a reallocation, which is prohibited on GPU kernels. This means that an array of a specific size has to be allocated first to contain all of the keys, but this is before the documents have been parsed and therefore the size required is unknown. A very large array could be allocated, but this would likely end up either wasting memory or still end up not being large enough. The push! command works outside of GPU kernels, but without a kernel we cannot fully exploit the speedup that a GPU provides.

Then, there is a question of updating duplicates. GPU CUDA Arrays in Julia cannot contain strings, so first each string would have to be converted into a pointer of the corresponding internalized string (pointers of internalized strings with the same contents yield the same pointers). Then, we would have to do a search through the array of previous pointers to see if there are any matches. One way to do this would be a linear scan, but even if we could parallelize this perfectly across the thousands of cores on the GPU, it is possible that we will be dealing with millions of different words, and the thousands of cores will not be enough to offset the additional $O(n)$ time complexity. Alternatively, we could maintain a sorted list and try to use binary search to make all inserts/searches take place in $\log n$ time, but this would not allow us to exploit the cores of a GPU well (since we cannot insert/search in parallel in the way that we would like since the operations would not be independent of each other and so could not be carried out separately on each core of the GPU).

Instead, we take the approach of processing each document on the CPU using dictionaries. On the CPU, we can use the push! command, and we can take advantage of the $O(1)$ amortized lookups that dictionaries provide. For each document, we create a dictionary counting all of the words in the document and their frequencies. Once this is done, we combine the dictionaries for each document into a single dictionary for all of the documents. 

Ultimately, the counts do have to be added, but this is where the GPU comes in. If there are a large number of long documents, as is typical in the MapReduce problem, then the number of keys (unique words in all of the documents) will likely be very large, and the array of counts for each key could have a length in the thousands. These counts could either be added on the CPU in parallel, or they could be transferred to the GPU for a quicker sum in parallel, inspired by the examples given in \cite{juliagpu} and \cite{rowsum}. 

We explore each of these possibilities in the following subsections. For sample documents, we used the inauguration speeches of Presidents Bush, Obama, Trump, and Biden, obtained from \cite{bushInauguration}, \cite{obamaInauguration}, \cite{trumpInauguration}, and \cite{bidenInauguration}. In order to best assess timings, we duplicated this set of $4$ speeches $n$ times for different values of $n$. We chose to duplicate these documents rather than finding large numbers of other documents because it allows for the simplest and clearest analysis of the timings of different functions in Section 4.1, where we summarize our findings for the different values of $n$.

The code for the following subsections can be found in the files titled ``CPU\_MapReduce.jl" and "GPU\_MapReduce.jl." The CPU\_MapReduce.jl file includes a MapReduce method that executes both Map and Reduce together on the CPU. Meanwhile, the GPU\_MapReduce.jl file includes a Map method that executes Map on the CPU and a block of code that represents the Reduce stage by calling a GPU kernel that quickly adds up the rows of matrices.

\subsection{Map on CPU}

As explained above, since dictionaries work on the CPU, we count the words in parallel on the CPU using the FLoops package and the command @floop, which is a generalization of typing ``Threads.@threads" before a for loop in Julia and allows the for loop to be executed in parallel. We use @floop because when we tried Threads.@threads or @sync @distributed, the code became much slower. However, @floop gave the desired effect of a parallel implementation.

We create a String array of the documents, and then iterate across that array. Then, we use the replace command to remove all punctuation (so that ``dog." and ``dog" count as the same word) and we use the split command to split each document into individual words.

Then, for each document, we initialize a dictionary, and we count the words in the document and their frequencies. We maintain an array of these dictionaries. After that, we initialize a large dictionary and we iterate in parallel through each of the dictionaries that were created for each document. Unlike the small dictionaries for each document, whose key value pairs were (string, count), the large dictionary has key value pairs (string, array of counts). This means that no addition is taking place in the map stage, but rather, the same keys are being grouped with each other for a fast reduce later on.

This code is implemented in the GPU\_MapReduce.jl file, since even though the reduce phase in that file is implemented on the GPU, the map phase still occurs on the CPU.


\subsection{Reduce on CPU}

For reducing on the CPU, we modify the code from mapping on the CPU ever so slightly. As described above, we add the counts together when populating the dictionary, ensuring that the CPU is doing the additions rather than outsourcing that to the GPU. In some ways, this is the most basic attempt at MapReduce, but we can still exploit parallelism on the CPU through the @floop macro just like in the map phase. This code can be found in the CPU\_MapReduce.jl file.


\subsection{Reduce on GPU}

We used the v100 GPU available on JuliaHub, which includes 8 vCPUs.

Another way is simply to make the big dictionary have the words as keys and an array of the counts from each of the documents that the word appeared in as values. For example, if we are parsing three documents, and the word ``the" appeared 10, 4, and 3 times respectively, then the big dictionary could either store (the, 17) or it could store (the, [10, 4, 3]).

In order to reduce on the GPU, we first need to convert the vectors of the counts for each key into a CuArray. In Julia, a vector of vectors cannot be converted into a CuArray, but a matrix can. Since not all of the vectors have the same lengths, we implement this by first initializing an array of zeros on the CPU that has as many rows as the number of keys and as many columns as the length of the longest vector of counts. Then, we populate the CPU array by filling in each row with the corresponding accounts. After that, we convert the array into a CUDA Array.

We also tried simply initializing a CUDA array of zeros and then populating each row with a CuArray of each of the vectors, but that took 3 times longer, so we do not pursue that approach. In the next section, we apply these algorithms and measure their timings and also see what the word frequencies are in Presidential speeches.


\ifx\allfiles\undefined

\bibliography{Index}

\end{document}

\fi

\newpage 

\ifx\allfiles\undefined

\documentclass[12pt,a4paper]{article}


\usepackage{graphicx}

\usepackage{amsmath}

\usepackage{amssymb}
\DeclareMathOperator{\tr}{trunk}
\usepackage{amsthm}

\usepackage{geometry}

\usepackage{fancyhdr}

\usepackage{color} 









\newcommand{\al}{\alpha}

\newcommand{\vphi}{\varphi}

\newcommand{\be}{\beta}

\newcommand{\ga}{\gamma}

\newcommand{\de}{\delta}

\newcommand{\om}{\omega}

\newcommand{\na}{\nabla}

\newcommand{\NA}{\nabla}

\newcommand{\bs}{\boldsymbol}

\newcommand{\ra}{\rightarrow}

\newcommand{\lra}{\longrightarrow}

\newcommand{\Ra}{\Rightarrow}

\newcommand{\xra}{\xrightarrow}

\newcommand{\xlra}{\xlongrightarrow}

\newcommand{\rgl}{\rangle}

\newcommand{\lgl}{\langle}

\newcommand{\dash}{\textrm{-}}

\newcommand{\ot}{\otimes}

\newcommand{\bpf}{\begin{proof}}

\newcommand{\epf}{\end{proof}}

\newcommand{\bthm}{\begin{thm}}

\newcommand{\ethm}{\end{thm}}

\newcommand{\bprop}{\begin{prop}}

\newcommand{\eprop}{\end{prop}}

\newcommand{\bcor}{\begin{cor}}

\newcommand{\ecor}{\end{cor}}

\newcommand{\blem}{\begin{lem}}

\newcommand{\elem}{\end{lem}}

\newcommand{\bdefn}{\begin{defn}}

\newcommand{\edefn}{\end{defn}}

\newcommand{\bexmp}{\begin{exmp}}

\newcommand{\eexmp}{\end{exmp}}

\newcommand{\brem}{\begin{rem}}

\newcommand{\erem}{\end{rem}}

\newcommand{\bdia}{\begin{displaymath}\xymatrix}

\newcommand{\edia}{\end{displaymath}}

\newcommand{\beq}{\begin{equation*}\begin{aligned}}

\newcommand{\eeq}{\end{aligned}\end{equation*}}

\newcommand{\bref}{\textbf{Ref}}

\newcommand{\intg}{\mathbb{Z}}

\newcommand{\real}{\mathbb{R}}

\newcommand{\comp}{\mathbb{C}}

\newcommand{\quot}{\mathbb{H}}

\newcommand{\afv}{\mathbb{A}}

\newcommand{\prv}{\mathbb{P}}

\newcommand{\mco}{\mathcal{O}}

\newcommand{\mcc}{\mathcal{C}}

\newcommand{\mcf}{\mathcal{F}}

\newcommand{\mcg}{\mathcal{G}}

\newcommand{\mcs}{\mathcal{S}}

\newcommand{\cp}{\mathbb{CP}}

\newcommand{\mfo}{\mathfrak{O}}

\newcommand{\mfg}{\mathfrak{g}}

\newcommand{\msa}{\mathscr{A}}

\newcommand{\msr}{\mathscr{R}}

\newcommand{\msg}{\mathscr{G}}

\newcommand{\msd}{\mathscr{D}}

\newcommand{\itbf}{\item\textbf}

\newcommand{\seqa}{a_1,...,a_}

\newcommand{\seqx}{x_1,...,x_}

\newcommand{\seqy}{y_1,...,y_}

\newcommand{\seqf}{f_1,...,f_}

\newcommand{\cred}{\textcolor{red}}

\newcommand{\cblue}{\textcolor{blue}}

\newcommand{\mfa}{\mathfrak{a}}

\newcommand{\mfb}{\mathfrak{b}}

\newcommand{\mfm}{\mathfrak{m}}

\newcommand{\mfn}{\mathfrak{n}}

\newcommand{\mfp}{\mathfrak{p}}

\newcommand{\Af}{A_{(f)}}


\newtheorem{thm}{\textbf {Theorem}}[section]

\newtheorem{cor}[thm]{\textbf{Corollary}}

\newtheorem{prop}[thm]{\textbf{Proposition}}

\newtheorem{lem}[thm]{\textbf{Lemma}}

\newtheorem{conj}[thm]{Conjecture}

\newtheorem{conv}[thm]{Convention}

\newtheorem{prob}[thm]{Problem}

\newtheorem{exer}[thm]{Exercise}

\newtheorem{quest}[thm]{Question}

\theoremstyle{definition}

\newtheorem{defn}[thm]{\textbf{Definition}}

\newtheorem{defns}[thm]{Definitions}

\newtheorem{exmp}[thm]{Example}

\newtheorem{exmps}[thm]{Examples}

\newtheorem{var}[thm]{Variant}

\newtheorem{vars}[thm]{Variants}

\newtheorem{con}[thm]{Construction}

\newtheorem{notn}[thm]{Notation}

\newtheorem{notns}[thm]{Notations}

\theoremstyle{remark}

\newtheorem{rem}[thm]{Remark}

\newtheorem{rems}[thm]{Remarks}

\newtheorem{warn}[thm]{Warning}

\newtheorem{sch}[thm]{Scholium}

\newtheorem{expl}[thm]{Explanations}

\newtheorem*{theorem}{\textbf{Theorem}}

\newtheorem*{corollary}{\textbf{Corollary}}

\newtheorem*{proposition}{\textbf{Proposition}}

\newtheorem*{lemma}{\textbf{Lemma}}

\newtheorem*{example}{\textbf{Example}}

\def\cok{\operatorname{Coker}}

\newcommand{\txi}{\tilde{\xi}}

\newcommand{\bxi}{\bar{\xi}}

\newcommand{\bz}{\bar{z}}

\begin{document}

\bibliographystyle{plain}

\else

\fi

\section{Results on Data}

\subsection{Timings and Analysis}

Here, we explore the timings for the CPU map phase and compare CPU vs GPU reduce timings. In the table, $n = 1$ corresponds to reading in 4 documents: the (first) inauguration speeches given by Presidents Bush, Obama, Trump, and Biden. Then, for larger values of $n$, we simply take $n$ copies of each of these 4 inauguration speeches (so $n = 100$ corresponds to counting the frequencies for $4n = 400$ documents, where there are $n$ duplicates of each speech).

The CPU we use here is the 32 thread machine on JuliaHub, while the GPU is the v100 on JuliaHub. All timings in the following subsections were calculated by calling @btime from the BenchmarkTools package in Julia.
Below is a table summarizing the results:

\begin{center}
\begin{tabular}{|c|c|c|c|}
    \hline
     $n$ & Map on CPU & Map \textbf{and} Reduce on CPU & Reduce on GPU \\
     \hline
     10 & 2.700 ms  & 3.971 ms & 444.406 $\mu$s \\ 
     \hline
     25 & 6.654 ms & 8.088 ms & 620.458 $\mu$s \\ 
     \hline
     50 & 11.258 ms & 16.651 ms & 1.152 ms \\ 
     \hline
     100 & 21.071 ms & 30.657 ms &  1.753 ms \\ 
     \hline
     250 & 49.808 ms &  82.518 ms & 7.256  ms \\ 
     \hline
     500 & 99.963 ms &  138.005 ms & 18.911  ms \\ 
     \hline
     
\end{tabular}
\end{center}

The graph below shows the time to do Map and Reduce on the CPU (third column above) vs the time to Map on CPU plus reduce on GPU (sum of second and fourth columns above) as $n$ gets larger:

\begin{figure}[H]
    \centering
    \fbox{\includegraphics[scale = 0.65]{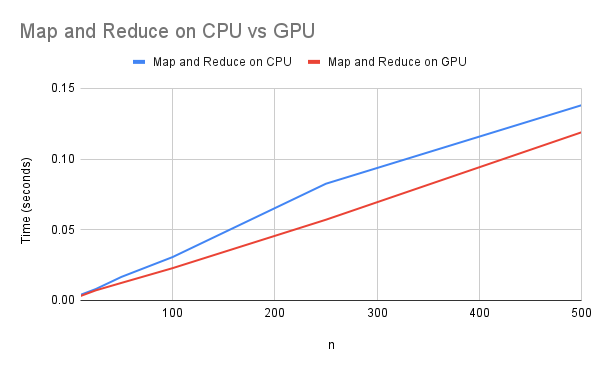}}
\end{figure}

We see that for small values of $n$ the speed up that the GPU offers for computing the additions is almost outweighed by the time it takes to initialize an array of zeros, populate it with the counts, and transfer that array to the GPU as a CuArray. However, this is not a concern because the MapReduce algorithm was created to work at scale on large numbers of documents at a time, so it is not relevant that it may not be as fast on the GPU when we are only dealing with small numbers of documents. In particular, we see that as $n$ reaches $250$ or $500$ (which corresponds to $1000$ or $2000$ documents), the GPU speedup is quite significant; in the case of $n = 500$, the GPU allows the MapReduce operation to occur 138.005 - 99.963 - 18.911 = 19.131 milliseconds faster.

However, despite the significant speedup that the GPU offers, we can see that the bulk of the time taken still occurs in the Map phase on the CPU. Running the Map phase on a machine with more than 32 threads available would allow the @floop parallelization to speed up the code more. Alternatively, this motivates the construction of an optimized dictionary on the GPU, an idea that we discuss further in Section 5.2: Future Research.

If we purely time the call to the CUDA kernel for the reduce phase, and we disregard the time it takes to initialize the array and transfer the memory to the GPU, we get a time of about 4.29 microseconds independent of the above values of $n$. Thus the main time sink for the GPU implementation is transferring the memory from the CPU to the GPU.

\subsection{Comparison to Julia's MapReduce Function}

Up until now, we have not studied Julia's own MapReduce implementations. There is one for the CPU, and the FoldsCUDA library has one for the GPU. In order to best compare our MapReduce implementations with Julia's own implementations, we compare them for finding the sum of the square roots of a set of positive numbers rather than for parsing documents. We focus on the optimized GPU implementation here rather than the CPU implementations.

For this reason, we make slight adjustments to our previous implementations of MapReduce to create a new implementation for this purpose. For the CPU implementation of map, we remove the loops for document parsing, and instead we simply add the numbers together using the @floop macro previously discussed. For the GPU implementation, we modify the row\_sum kernel function that we discussed previously so that it sums square roots of numbers. The reason for mapping $x \to \sqrt{x}$ was because mapping $x \to x$ or $x \to x^2$ results in the compiler optimizing away the loop, thereby completing the computation independent of $n$ in approximately one nanosecond. On the other hand, there is no simple closed form for adding up square roots so this is a better way to compare the timings of the different algorithms.

In order to time Julia's GPU implementation, we simply called @btime FoldsCUDA.mapreduce($\sqrt{x}$, +, 1:n). This code can be found in the MapReduceAdd.jl file. The computations below represent the time required  to compute $\sum_{i = 1}^n \sqrt{i}$ for each value of $n$ listed in the ways described above.

\begin{center}
\begin{tabular}{|c|c|c|}
    \hline
     $n$ & Our GPU Sum & Julia GPU Sum \\
     \hline
     10  & 3.494 $\mu$s & 19.843 ns\\ 
     \hline
     $10^2$  & 4.532 $\mu$s & 263.509 ns \\ 
     \hline
     $10^3$ & 4.311 $\mu$s & 2.379 $\mu$s \\ 
     \hline
     $10^4$  & 4.466 $\mu$s & 27.260 $\mu$s\\ 
     \hline
     $10^5$ & 4.406 $\mu$s & 278.826 $\mu$s \\ 
     \hline
     $10^6$  & 4.544 $\mu$s & 2.794 ms \\ 
     \hline
     $10^7$ & 4.499 $\mu$s & 28.017 ms \\ 
     \hline
     $10^8$  & 73.372 ms & 280.168 ms \\ 
     \hline
     
\end{tabular}
\end{center}

We can also visualize this graphically below:

\begin{figure}[H]
    \centering
    \fbox{\includegraphics[scale = 0.65]{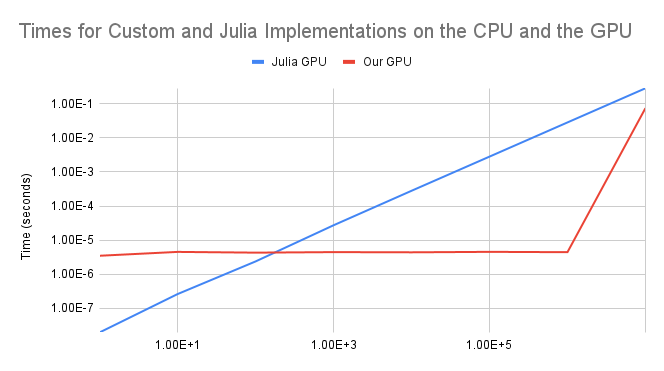}}
    \label{fig:my_label}
\end{figure}

We can see from this table and graph that our GPU implementation is much slower for $n < 1000,$ but once $n \geq 10000,$  our kernel is faster than Julia's FoldsCUDA.MapReduce package. This is only one specific example, and The FoldsCUDA MapReduce implementation is much more general. It is also worthwhile noting that our GPU implementation is very flat up until $n = 10^8$ where it spikes dramatically, and the spike is clearly visible even with a log scale on the y axis. Meanwhile, the FoldsCUDA timing increases linearly with on this log log plot, which indicates that the timing increases linearly on a non scaled plot as well. However, this still shows that a relatively simple GPU kernel can outperform the GPU MapReduce for this specific type of problem. This kernel is using atomic add to ensure that the answer is correct; this could likely be optimized further, but it still beats the FoldsCUDA implementation.

In addition to doing $\sum_{i =1}^n \sqrt{i}$ we can also try computing $\sum_{i = 1}^n \frac{1}{i} \cdot (-1)^{i + 1}$ through our GPU implementation and the FoldsCUDA implementation. This works because the compiler will not be able to optimize the computation away. From calculus, we know that this alternating harmonic sum converges to $\ln 2 \approx 0.693147.$ If we try this, we get a similar pattern in the timings:

\begin{center}
\begin{tabular}{|c|c|c|}
    \hline
     $n$ & Our GPU Sum & Julia GPU Sum \\
     \hline
     10  & 6.561 $\mu$s & 62.372 ns\\ 
     \hline
     $10^2$  & 6.497 $\mu$s & 814.608 ns \\ 
     \hline
     $10^3$ & 6.608 $\mu$s & 29.387 $\mu$s \\ 
     \hline
     $10^4$  & 6.505 $\mu$s & 485.004 $\mu$s\\ 
     \hline
     $10^5$ & 6.649 $\mu$s & 5.222 ms \\ 
     \hline
     $10^6$  & 6.896 $\mu$s & 54.927 ms \\ 
     \hline
     $10^7$ & 6.525 $\mu$s & 626.515 ms \\ 
     \hline
     $10^8$  & 65.323 ms & 6.515 s \\ 
     \hline
     
\end{tabular}
\end{center}

Curiously, we once again observe the relative consistency of our GPU sum followed by a sudden spike for $n = 10^8.$ However, even with $n = 10^8$, our implementation is still about 100 times faster than Julia's FoldsCUDA implementation for this problem. In the next section, we will return to reading documents, and we will apply our word frequency MapReduce algorithms to detect patterns in Presidential speeches and what words they use.

\subsection{Results on Presidential Speeches}

In the previous subsections, we described the times for running the GPU algorithm for MapReduce on the CPU and GPU. We saw that the GPU gives a slight speedup, particularly when there are more documents and they are longer. Here, we focus less on the speed of implementation (since as we saw, it is only a matter of microseconds) and more on the results from running the algorithm to see if we can discern anything from the word frequencies on specific documents.

There are a myriad of possible documents that could be studied. Here, we analyze Presidential speeches from Presidents George W Bush, Barack Obama, Donald Trump, and Joseph Biden, specifically their State of the Union speeches (retrieved from \cite{bidenSOTU}, \cite{bushSOTU}, \cite{obamaSOTU}, and \cite{trumpSOTU}) and Inauguration Speeches. All public Presidential speeches are legally required to be made available to the public by the Presidential Records Act, allowing for easy access to them. These documents are included in the folders containing speeches for each President. Many Presidents use similar words, from discussing unity and patriotism to more common words like ``the" or ``our." But are there more unique words that each US President tends to use that distinguishes them (or perhaps, their speechwriters) from their fellow Presidents?

It turns out the answer is yes. President Biden is easily identifiable by his use of the word ``troops." While every commander in chief speaks about the US military. President Biden is the only one who ends every one of his speeches with ``And may God protect our troops." The other words in this sentence are frequently used by other Presidents and native English speakers very frequently (especially words like ``our" and ``and"), so they are ineffective in helping identify which President gave the speech.

President Trump stands out from other Presidents from his use of the word ``God." While every President invokes God from time to time, President Trump mentioned God an average of 5 times per speech in the speeches sampled, compared to 1 or 2 times per speech for the other Presidents. This is not surprising though, as President Trump's base included many religious evangelical Christians who would be very supportive of their President invoking God.

President Obama stands out for his use of the word ``Iran," likely in reference to Foreign Policy and the 2015 Iran Nuclear Deal. Across these speeches, President Obama used the word ``Iran" 12 times, while President Bush used it 3 times, President Trump used it 5 times, and President Biden has not used it at all.

President Bush stands out by his use of the word ``terrorist" in his State of the Union addresses and inauguration speeches, likely in reference to the September 11th, 2001 attacks. He used this word an average of 4 times per speech, while his successors either never used the word ``terrorist" or at most once or twice.

Another interesting trend is to look at the number of times each President received an applause, which is marked by the word ``(Applause)" appearing in the speech transcript. In the speeches sampled, President Bush received an applause 220 times, President Obama received an applause 210 times, President Trump received an applause 131 times, and President Biden received an applause only 17 times. The fact that President Biden received so few is likely because his speeches did not consist of as many State of the Union addresses as his predecessors since he has not given many SOTU addresses yet. Nevertheless, the fact that the number of applauses has decreased since President Bush regardless of party may suggest further political polarization, where even applauding for the President of the United States is perhaps not as unifying as it used to be during President Bush's term in the wake of 9/11.

Because there are so many words that all Presidents and speakers use, it is not practical to code an ML method like a neural network to analyze the most common words and predict which President said them. These observations were made by considering the political climate of the time and confirming that the President's language in his speeches matched what we would expect given the time period.

\newpage 

\ifx\allfiles\undefined

\documentclass[12pt,a4paper]{article}


\usepackage{graphicx}

\usepackage{amsmath}

\usepackage{amssymb}
\DeclareMathOperator{\tr}{trunk}
\usepackage{amsthm}

\usepackage{geometry}

\usepackage{fancyhdr}

\usepackage{color} 









\newcommand{\al}{\alpha}

\newcommand{\vphi}{\varphi}

\newcommand{\be}{\beta}

\newcommand{\ga}{\gamma}

\newcommand{\de}{\delta}

\newcommand{\om}{\omega}

\newcommand{\na}{\nabla}

\newcommand{\NA}{\nabla}

\newcommand{\bs}{\boldsymbol}

\newcommand{\ra}{\rightarrow}

\newcommand{\lra}{\longrightarrow}

\newcommand{\Ra}{\Rightarrow}

\newcommand{\xra}{\xrightarrow}

\newcommand{\xlra}{\xlongrightarrow}

\newcommand{\rgl}{\rangle}

\newcommand{\lgl}{\langle}

\newcommand{\dash}{\textrm{-}}

\newcommand{\ot}{\otimes}

\newcommand{\bpf}{\begin{proof}}

\newcommand{\epf}{\end{proof}}

\newcommand{\bthm}{\begin{thm}}

\newcommand{\ethm}{\end{thm}}

\newcommand{\bprop}{\begin{prop}}

\newcommand{\eprop}{\end{prop}}

\newcommand{\bcor}{\begin{cor}}

\newcommand{\ecor}{\end{cor}}

\newcommand{\blem}{\begin{lem}}

\newcommand{\elem}{\end{lem}}

\newcommand{\bdefn}{\begin{defn}}

\newcommand{\edefn}{\end{defn}}

\newcommand{\bexmp}{\begin{exmp}}

\newcommand{\eexmp}{\end{exmp}}

\newcommand{\brem}{\begin{rem}}

\newcommand{\erem}{\end{rem}}

\newcommand{\bdia}{\begin{displaymath}\xymatrix}

\newcommand{\edia}{\end{displaymath}}

\newcommand{\beq}{\begin{equation*}\begin{aligned}}

\newcommand{\eeq}{\end{aligned}\end{equation*}}

\newcommand{\bref}{\textbf{Ref}}

\newcommand{\intg}{\mathbb{Z}}

\newcommand{\real}{\mathbb{R}}

\newcommand{\comp}{\mathbb{C}}

\newcommand{\quot}{\mathbb{H}}

\newcommand{\afv}{\mathbb{A}}

\newcommand{\prv}{\mathbb{P}}

\newcommand{\mco}{\mathcal{O}}

\newcommand{\mcc}{\mathcal{C}}

\newcommand{\mcf}{\mathcal{F}}

\newcommand{\mcg}{\mathcal{G}}

\newcommand{\mcs}{\mathcal{S}}

\newcommand{\cp}{\mathbb{CP}}

\newcommand{\mfo}{\mathfrak{O}}

\newcommand{\mfg}{\mathfrak{g}}

\newcommand{\msa}{\mathscr{A}}

\newcommand{\msr}{\mathscr{R}}

\newcommand{\msg}{\mathscr{G}}

\newcommand{\msd}{\mathscr{D}}

\newcommand{\itbf}{\item\textbf}

\newcommand{\seqa}{a_1,...,a_}

\newcommand{\seqx}{x_1,...,x_}

\newcommand{\seqy}{y_1,...,y_}

\newcommand{\seqf}{f_1,...,f_}

\newcommand{\cred}{\textcolor{red}}

\newcommand{\cblue}{\textcolor{blue}}

\newcommand{\mfa}{\mathfrak{a}}

\newcommand{\mfb}{\mathfrak{b}}

\newcommand{\mfm}{\mathfrak{m}}

\newcommand{\mfn}{\mathfrak{n}}

\newcommand{\mfp}{\mathfrak{p}}

\newcommand{\Af}{A_{(f)}}


\newtheorem{thm}{\textbf {Theorem}}[section]

\newtheorem{cor}[thm]{\textbf{Corollary}}

\newtheorem{prop}[thm]{\textbf{Proposition}}

\newtheorem{lem}[thm]{\textbf{Lemma}}

\newtheorem{conj}[thm]{Conjecture}

\newtheorem{conv}[thm]{Convention}

\newtheorem{prob}[thm]{Problem}

\newtheorem{exer}[thm]{Exercise}

\newtheorem{quest}[thm]{Question}

\theoremstyle{definition}

\newtheorem{defn}[thm]{\textbf{Definition}}

\newtheorem{defns}[thm]{Definitions}

\newtheorem{exmp}[thm]{Example}

\newtheorem{exmps}[thm]{Examples}

\newtheorem{var}[thm]{Variant}

\newtheorem{vars}[thm]{Variants}

\newtheorem{con}[thm]{Construction}

\newtheorem{notn}[thm]{Notation}

\newtheorem{notns}[thm]{Notations}

\theoremstyle{remark}

\newtheorem{rem}[thm]{Remark}

\newtheorem{rems}[thm]{Remarks}

\newtheorem{warn}[thm]{Warning}

\newtheorem{sch}[thm]{Scholium}

\newtheorem{expl}[thm]{Explanations}

\newtheorem*{theorem}{\textbf{Theorem}}

\newtheorem*{corollary}{\textbf{Corollary}}

\newtheorem*{proposition}{\textbf{Proposition}}

\newtheorem*{lemma}{\textbf{Lemma}}

\newtheorem*{example}{\textbf{Example}}

\def\cok{\operatorname{Coker}}

\newcommand{\txi}{\tilde{\xi}}

\newcommand{\bxi}{\bar{\xi}}

\newcommand{\bz}{\bar{z}}

\begin{document}

\bibliographystyle{plain}

\else

\fi

\section{Conclusion}

\subsection{Summary}

We began with an MPI implementation, which aimed to parallelize both the map and reduce stages, but we found that there were some issues with fully reducing the key value pairs. Then, we switched to a GPU implementation, where we found that the map phase is best done on the CPU but the counting/reduce phase was optimally done on the GPU. We also saw that the GPU reduce implementations of $\sum_{i = 1}^n \sqrt{i}$ and $\sum_{i = 1}^n (-1)^{i + 1} \frac{1}{i}$ are significantly faster than the FoldsCUDA implementation on this problem, highlighting the potential for future improvement of a faster GPU MapReduce algorithm in general.

After that, we switched to applying our MapReduce algorithm to count the frequencies of words in US Presidential speeches, and we found that while the Presidents used many common words, there were also some words that were different among them, a result of the political climate of the time and the Presidents themselves.

\subsection{Future Directions of Study}

This paper revealed several possible approaches to implement MapReduce in parallel. One way to further the research in this field would be to combine MPI to perform map reduce across multiple GPUs, adding another layer of parallelism for particularly large computations. Another would be to first implement an optimized dictionary on GPUs, and then perform most if not all of the GPU algorithm exclusively on the GPU, rather than parsing the documents with dictionaries on the CPU. 

A dictionary available on the GPUs would likely optimize both the map and reduce phases: the map phase could be performed in parallel across the GPU's thousands of cores, and there would not need to be a memory transfer from the CPU to the GPU for the reduce phase to take place, cutting the reduce phase for the large numbers of documents that we saw in Section 3 from milliseconds to microseconds.

The algorithm can also be applied to more data. Here, we focused on counting words in Presidential speeches, which is only one of countless possible applications of counting the frequencies in documents. Another set of documents to study would be transcripts of ordinary dialogue of native speakers in different languages to compare what words are often used, and whether they are the same up to translation across languages or if there are specific wors used more in certain languages than others.

In addition, there are many applications where it is crucial to add up a large number of numbers, and this ties into the MapReduce algorithm we discussed in Section 3.5 about computing $\sum_{i = 1}^n \sqrt{i}$ and $\sum_{i = 1}^n (-i)^{i + 1} \frac{1}{i}$ for large $n$. We only used a fairly simple GPU algorithm to optimize that computation, but that GPU kernel could almost certainly be optimized significantly more and that would allow a large number of MapReduce applications to be executed faster.

\section*{Acknowledgments}

I would like to thank the MIT Computer Science and Artificial Intelligence Laboratory, the MIT Math Department, Professor Alan Edelman and TA Ranjan Anantharaman for hosting the course 18.337/6.338, allowing me to conduct this project, and giving me feedback on my proposal. I also thank the MIT Julia lab for providing me with JuliaHub computing access to run the MPI and GPU simulations described in this paper. Lastly, I thank the members of the Julia Slack and the gpu channel for providing feedback on the ideas discussed in this project.


\ifx\allfiles\undefined

\bibliography{Index}

\end{document}

\fi

\ifx\allfiles\undefined

\bibliography{Index}

\end{document}

\fi


\ifx\allfiles\undefined

\bibliography{Index}

\end{document}

\fi

\bibliography{Index}

\end{document}